\documentclass[12pt,nonbibnotes,nofootinbib,notitlepage]{revtex4}
\usepackage{amsmath}
\usepackage{amssymb}
\usepackage{braket}
\usepackage{graphicx}

\usepackage[cmyk]{xcolor}
\usepackage[unicode, colorlinks, urlcolor=blue, citecolor=blue, linkcolor=blue]{hyperref}
\usepackage[all]{hypcap}
\usepackage[T2A]{fontenc}
\usepackage[cp1251]{inputenc}
\usepackage[english,russian]{babel}
\usepackage{cmap}

\begin{document}
\title{Exciton-polariton interference controlled by electric field
}

\author{D. K. Loginov, P. A. Belov, V. G. Davydov, I. Ya. Gerlovin, I. V. Ignatiev, A. V. Kavokin}
\affiliation{Spin Optics Laboratory, St. Petersburg State University, Ulyanovskaya 1, Petrodvorets, 198504 St. Petersburg, Russia}

\author{Y. Masumoto}
\affiliation{Institute of Physics, Tsukuba University, Tsukuba, 305-8571, Japan}

\date{\today}

\begin{abstract}
Linear in the wave-vector terms of an electron Hamiltonian play an important role in topological insulators and spintronic devices. Here we demonstrate how an external electric field controls the magnitude of a linear-in-K term in the exciton Hamiltonian in wide GaAs quantum wells. The dependence of this term on the applied field in a high quality sample was studied by means of the differential reflection spectroscopy. An excellent agreement between the experimental data and the results of calculations using semi-classical non-local dielectric response model confirms the validity of the method and paves the way for the realisation of excitonic Datta-and-Das transistors. In full analogy with the spin-orbit transistor proposed by Datta and Das [Appl. Phys. Lett. {\bf 56}, 665 (1990)], the switch between positive and negative interference of exciton polaritons propagating forward and backward in a GaAs film is achieved by application of an electric field with non-zero component in the plane of the quantum well layer.
\end{abstract}

\maketitle

\section{Introduction}

GaAs is the best studied direct band gap semiconductor nowadays. A remarkable progress of the epitaxial growth technology made it possible to fabricate nearly ideal layers of GaAs, where very fine quantum effects may be studied. Engineering of quantum properties of quasiparticles in semiconductor heterostructures has led to a set of remarkable discoveries in the past decades. Spin-orbit interaction in such structures drastically changes the dispersion of electron moving in a crystal and leads to appearance of the effective magnetic field, strongly related to the wave vector of the carrier, and affecting the electron spin ~\cite{Manchon,GlazovBook}. These fundamental properties have been originally understood by Dresselhaus and Rashba as coupling of the odd in the electron’s wave vector $k$ for crystals with inversion symmetry breaking~\cite{Dresselhaus, Rashba, PikusMaruschak}. 

Linear-in-$k$ terms of the electronic Hamiltonian are in the heart of the spin Hall effect~\cite{Hirsch, Kato, Wunderlich} and the Datta-and-Das~\cite{DattaDas} transistor proposal. They play a key role in topological insulators~\cite{Klembt2018} and are essential for the research on Majorana fermions~\cite{LutchynPRL2010,Stevan2014,He2017}. While the linear-in-$k$ terms imposed by spin-orbit coupling in semiconductors are extensively studied by now~\cite{PikusMaruschak, IvchenkoPikus, GlazovBook, Luo}, less is known about spin-independent linear-in-$k$ terms that may be induced by strain or external electric field. From the group theory point of view, the existence of such terms is straightforward, however their experimental detection in transport or optical measurements is still challenging. 

Here we show that a good tool for the characterization of spin-independent linear-in-$K$ terms may be offered by the differential reflectivity spectroscopy of semiconductor quantum wells. The signatures of quantum confined states of light-matter quasiparticles, exciton-polaritons, in reflectivity spectra contain a precious information on the exciton kinetic Hamiltonian. We have found that varying the magnitude of the linear-in-$K$ terms by tuning the external electric field, one should be able to invert the shape of excitonic resonances in the reflectivity spectra. The presence of the phase-inverted resonances is the irrefutable evidence for the linear-in-$K$ terms, while the dependence of the shape of the resonances on the applied field provides a quantitative information on the magnitude of these terms. Moreover, the switching between positive and negative interference of exciton polariton modes propagating in forward and backward directions induced by the in-plane electric field constitutes a clear manifestation of the Datta-and-Das effect for excitons.

Several previous studies were devoted to the excitonic effects induced by linear in K terms. In particular, a $K$-dependence of the exciton $g$-factor~\cite{Koch-PRL2006, Koch-PRB2011, Grigoryev-PRB2016}, an increase of the exciton-light coupling in the longitudinal magnetic field directed along the heterostructure growth axis~\cite{Grigoryev-PRB2017}, and an increase of the exciton effective mass in a transverse magnetic field~\cite{Bodnar-PRB2017} are reported. 
In Refs.~\cite{Loginov-PRB2014, Loginov-pss2016}, an effect of an uniaxial strain on the exciton states is experimentally and theoretically studied for wide QWs. A nontrivial effect of the phase inversion of the spectral oscillations has been predicted. It is related to the appearance of the linear-in-$K$ terms in the exciton Hamiltonian in the presence of strain.

The effect of an electric field on the exciton states in semiconductor crystals and heterostructures has been studied already for several decades~\cite{Gross1962, Frova1966, Blossey1970, Monozon-PRB2010, Raczynska-2016, Glazov-2017, Glazov-2018,Ivchenko-book, Kavokin-book2017,Baldo-2009, Cohen-2011}. The works studying effects of the electric field on excitons mainly devoted to the relatively narrow QWs. The application of an electric field gives rise in this case to a modification of the relative electron-hole motion in an exciton~\cite{Ivchenko-book, Kavokin-book2017} thus causing the Stark shift of exciton states~\cite{Bassani-2000, Kuo-2005, Zhang-2009} and inducing a static dipole moment of excitons~\cite{Butov-2017, Savvidis-PRL2018}. 

The wide QWs, however, are of a particular interest because they offer an opportunity to study the effect of an electric field on the motion of the exciton as a whole particle. The excitons with large wave vectors, $K \gg q$, where $q$ is the wave vector of light, can be experimentally observed due to their quantum confinement in the wide QWs. The wave vector selection rules for optical transitions are broken by the QW interfaces. The model of quantization of the exciton motion across the QW layer is well verified for the wide QWs~\cite{Kiselev-pss1975, Ivchenko-book, Khramtsov-JAP2016, Khramtsov-PRB2019}. In contemporary optical experiments with wide QWs in high-quality heterostructures one can observe many quantum-confined exciton states. The states are typically observed as resonant features (oscillations) in the reflectance spectra of the heterostructures~\cite{Kiselev-pss1975, Tredicucci-PRB1993, Tomasini-PRB1995, Ubyivovk-JLumin2003, AndreaCho-PRB2004, Loginov2006, Nakayama-PRB2006, Trifonov-PRB2015, Grigoryev-SuperMicro2016, Khramtsov-PRB2019}. This allows one to study the effects of the external fields on the propagating excitons.

There are very few studies of the electric field effects for excitons with large $K$ vector. One can refer to the paper by Zielinska-Raczynska {\it et al.}~\cite{Raczynska-2018}, where an electro-optical function for $P$-excitons in a thin Cu$_2$O plate is theoretically analyzed. However, only a particular case of the co-directed electric field and the exciton $K$-vector, which are perpendicular to the plate surface, has been considered.  These effects are studied in relatively small electric fields where the exciton ionization is not important. In the QW structures, a relatively strong electric field can be applied across the QW layer where the ionization processes are effectively blocked by the barriers. Application of a strong electric field along the QW layer is impractical because of various secondary effects related to the presence of resident or photocreated carriers. They are accelerated by the electric field and destroy the exciton states.

In this work we study the quantum-confined exciton states in a wide QW in the presence of an electric field, which contains a non-zero in-plane component. The electroreflectance spectra were measured for a heterostructure with the 120-nm GaAs QW. Multiple resonant peculiarities (spectral oscillations) related to the quantum-confined exciton states are observed in the spectra. The application of an electric field $F$, tilted at a small angle to the structure growth axis, is found to reduce the oscillations almost down to zero amplitude at some critical value of the electric field $F_c$. However the oscillations appear again as the field increases beyond this critical value. The phase of these oscillatory features in the spectra becomes inverted with respect to that for the case of $F < F_c$.
In other words, the electric field can be used to control the phase of the spectral resonances. It opens up an opportunity to design the polariton interferometry devices similar to those based on the electro-optical effect~\cite{DattaDas, Sheremet2016}.

A theoretical model accounting for the interference of polaritonic waves is developed to describe the spectra. The inversion of the phase of spectral oscillations is shown to be governed by the linear in exciton wave vector $K$ term of the exciton Hamiltonian that is induced by the in-plane electric field component. The electroreflectance spectra simulated in the framework of a polaritonic model well reproduce the main features observed in the experiment. The only free parameter of the model is the factor $\lambda$ that defines the magnitude of linear in $K$ term. An estimate of this factor is also given.

\section{Experiment}
\label{sec:experiment}
The studied semiconductor structure was grown on a n-doped GaAs (001) substrate by the gas source molecular beam epitaxy. It contains several heterolayers including QWs, superlattices and quantum dots. Here we consider only a thick GaAs QW, which was grown between a short-period GaAs/AlAs superlattice and a thin AlAs layer followed by a  thick In$_{0.51}$Ga$_{0.49}$P barrier layer. The nominal width of the QW is 120~nm. 

The sample was provided with a semitransparent indium tin oxide (ITO) electrode (Schottky contact) on the top surface and with an Ohmic contact on the back surface. Both the applied bias, $U_{\mathrm{bias}}$, and the electric current flowing through the sample were controlled during the experiments. Measurements have shown that the current exponentially increases as the bias is approaching to +1~V. 
The absolute value of the current also rises superlinearly at negative bias $U_{\mathrm{bias}} < -2$~V.

Reflectrance spectra were measured using a tunable continuous-wave titan-sapphire laser as a light source. The laser beam was directed almost perpendicularly to the sample surface. It was focused in a relatively small spot with a diameter of about 100~$\mu$m near the edge of the electrode. The bias was applied to a point contact at the opposite edge of the 5-mm round electrode. Due to the relatively large resistance of this thin electrode of about 3~k$\Omega/$cm, the voltage applied to the sample in the illuminated area deviated from that applied under the contact. What is important for the present work, the electric field created by the voltage was directed at some angle to the growth axis $z$, that is it has a nonzero in-plane component $F_x$. As we discuss below, this component is responsible for the observed effect. To increase the detection sensitivity of the electric-field-induced variations of the reflectance spectra, an alternative-current (AC) voltage with the amplitude of 0.1~V at the frequency of 100~kHz was applied in addition to the bias. Besides, the intensity of laser beam was modulated at the frequency of 2~kHz. A double lock-in detection of the signal modulated at both frequencies allowed us to detect small changes of reflection as low as 10$^{-6}$. In fact, the electro-reflectance spectra were measured in our experiments~\cite{Pollak-PR1968}.

\begin{figure}[!h]
\includegraphics[width=0.65\textwidth]{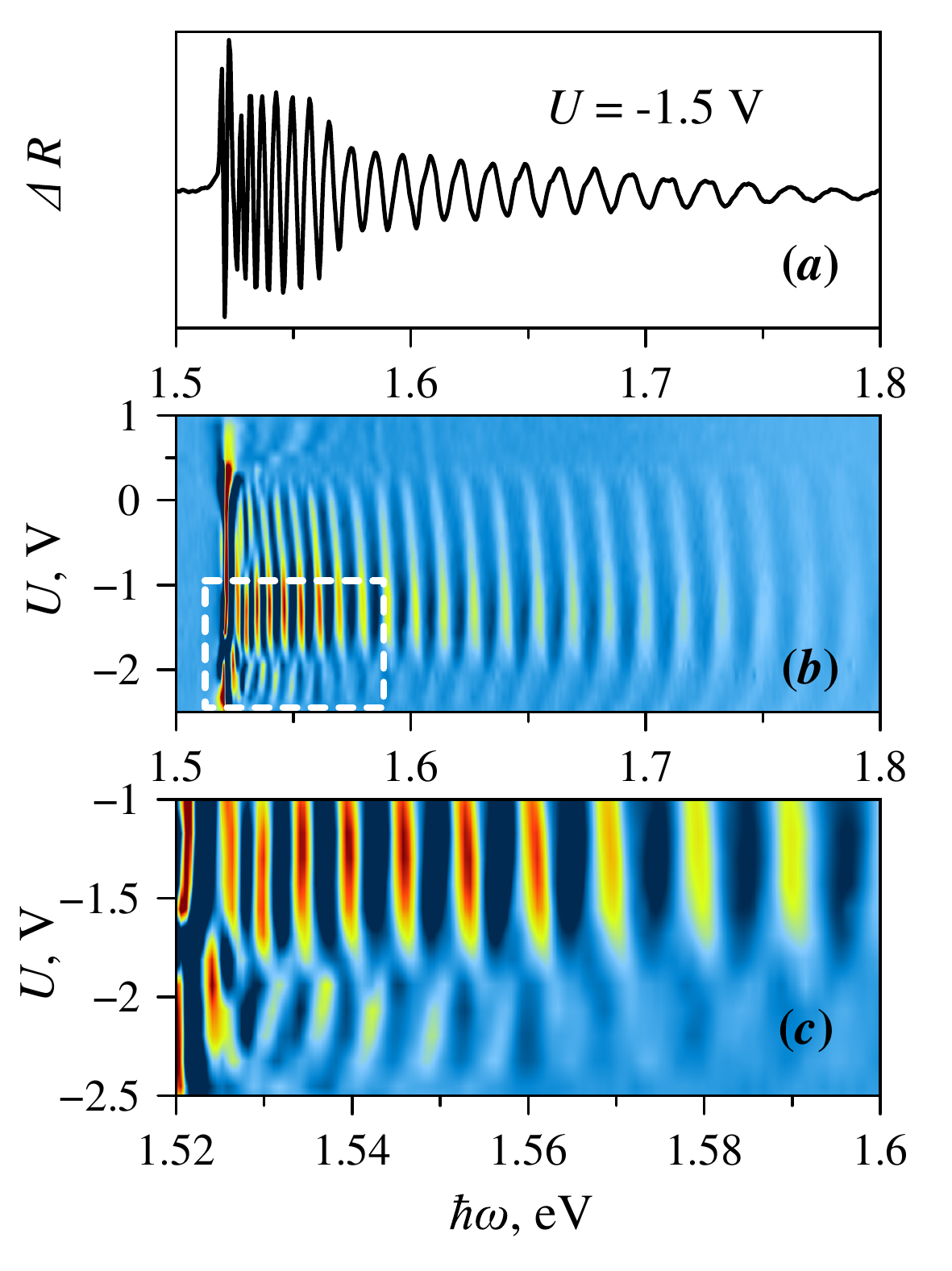}
\caption{(a) An example of the electroreflectance spectrum of a heterostructure with the 120-nm GaAs QW at $U_{\textrm{bias}} = -1.5$~V. (b) Two-dimensional plot illustrating evolution of the spectra with the bias varied from +1~V to -2.5~V. (c) Zoom to (a) showing the effect of the phase inversion of the spectral oscillations.
}
\label{Fig1}
\end{figure}

An example of an electro-reflectance spectrum measured at $U_{\textrm{bias}} = -1.5$~V is shown in Fig.~\ref{Fig1}(a). Many oscillations in the wide spectral range 1.5 -- 1.8~eV are observed in the spectrum. They can be attributed to the quantum-confined exciton states in the QW. Similar oscillations were measured in reflectance spectra of heterostructures with wide QWs by many authors~\cite{Tredicucci-PRB1993, Tomasini-PRB1995, Ubyivovk-JLumin2003, Koch-PRB2011, Khramtsov-PRB2019}. However, such oscillations are typically observed in a relatively narrow spectral range, of order of several tens of meV. Here, due to the strongly increased sensitivity, we managed to measure the oscillations in the broader spectral range.

The energy distance between the oscillations gradually increases. An analysis of the experimental data gives rise to a phenomenological dependence for the oscillation energies: $E_n \approx E_g + a \cdot n^{5/3}$ where $E_g = 1.52$~eV is the band gap in GaAs and $n$ is the oscillation number. It is well known that quantization of the exciton motion as a single particle leads to quadratic dependence of the energy on its number~\cite{Tredicucci-PRB1993, Ubyivovk-JLumin2003}. This dependence comes from the assumed parabolic dispersion for the electron and hole energy states. However, for the broad energy range, the dispersion, in particular the electron dispersion, deviates from the parabolic law~\cite{IvchenkoPikus}. It explains the observed energy dependence of the oscillations. We take into account this dependence in our theoretical modeling below.

The application of voltage in the range $U_{\textrm{bias}} = 0 \ldots -1.5$~V does not considerably affect the spectral oscillations, see Fig.~\ref{Fig1}(b). Only small energy shifts and variations of the oscillation amplitudes are observed at these values of bias. However, further increase of the bias (in absolute value) is followed by a dramatic decrease of the oscillation amplitude. Such an effect would be expected due to fast ionization of excitons in the electric field with a subpicosecond characteristic time. This is, however, not the case for the experimental data under discussion. Indeed, the spectral oscillations appear again at $U_{\textrm{bias}} < -2$~V. The phase of the appearing oscillations is opposite to that for small bias, i.e., the spectral maxima and dips are inverted at the strong bias. This effect is illustrated in detail in Fig.~\ref{Fig1}(c). 

A similar effect of the phase inversion of spectral oscillations related to the quantum-confined exciton states has been theoretically predicted for reflectance spectra of the heterostructures with wide QWs uniaxially compressed perpendicularly to the growth axis~\cite{Loginov-PRB2014}. As it was found in that work, the phase inversion is a result of the linear-in-$K$ term appearing in the exciton Hamiltonian of the strained QW structure. We may assume, therefore, that the phase inversion observed in the tilted electric field has a similar origin. The theoretical analysis described below confirms this assumption.

\section{Theory}
\label{sec:theory}

It is well known that spectral oscillations of the reflection coefficient can be described as an interference of the forward and backward polaritonic waves propagating in a wide QW~\cite{YChen, Tomasini-PRB1995, Loginov2006, Ivchenko-book, Loginov-PRB2014, Kavokin-book2017}. The phase inversion observed experimentally at the bias of $U_{\textrm{bias}} \approx -1.8$~V can, therefore, be a result of the electric-field-induced phase shift between these waves. Similarly to work~\cite{Loginov-PRB2014}, we assume that this phase shift is due to the linear-in-$K$ term in the exciton Hamiltonian. As we show below, this term appears due to the modification of the wave function of the relative electron-hole motion by the in-plane electric field component.

According to the model accounting for the interference of the polaritonic waves, a reflectance spectrum is determined by a dielectric function depending on the polariton wave vector (the effect of spatial dispersion)~\cite{Ivchenko-book, Kavokin-book2017}. Following the experimental conditions, we assume that the incident linearly polarized light-wave propagates perpendicularly to the heterostructure surface. The linear polarization is a superposition of two circular components creating excitons with the angular momentum projections $S_z=+1$ and $S_z=-1$. We consider the dielectric function for one component only. The expression for an other component is similar. 

The dielectric function, which takes into account the exciton-light interaction, is described by expression~\cite{Ivchenko-book, Kavokin-book2017,Tomasini-PRB1995, Loginov2006}:
\begin{equation}
\varepsilon(\omega, K)=\varepsilon_0+\frac{\varepsilon_0\hbar\omega_{LT}(F)}{H(F, K)-\hbar\omega+i\hbar\Gamma}.
\label{Eq1}
\end{equation}
Here, $\varepsilon_0$ is the background dielectric constant, $\hbar\omega_{LT}$ is the longitudinal-transverse exciton splitting describing the exciton-light interaction strength, $\omega$ is the frequency of light, $\hbar\Gamma$ is the exciton damping rate. Term $H(F, K)$ is the exciton energy depending on the electric field and the exciton wave vector. This energy is obtained from the exciton Hamiltonian, whose dependence on the electric field eventually determines behavior of the reflectance spectra.

\subsection{The exciton Hamiltonian in the presence of electric field}
\label{sec:hamiltonian}

Let us consider an exciton moving along the $z$-axis coinciding with the [001] crystal axis and the growth axis of the wide QW made of a semiconductor crystal having a zinc-blende symmetry. Axes $x$ and $y$ are assumed to be directed along the [100] and [010] crystal axes, i.e. along the QW plane. The corresponding components of the exciton wave vector are taken to be $K=K_z$, $K_x=K_y=0$.

The exciton Hamiltonian includes the Hamiltonian of a free electron, the Luttinger Hamiltonian of a free hole~\cite{IvchenkoPikus}, the term describing the Coulomb electron-hole interaction, and the term describing the effect of an electric field. We consider hereafter only the heavy-hole excitons because their contribution into the reflectance spectra is considerably larger than that of the light-hole excitons~\cite{Loginov2006, Khramtsov-PRB2019}. 

Since the QW width is much larger than the exciton Bohr radius, we shall use the model of quantization of the exciton center-of-mass motion. Rearranging the terms of the exciton Hamiltonian, we can represent the energy operator as
\begin{equation}
\hat H=E_g+\hat H_K+\hat H_{p} + \hat V,
\label{eq:Hamiltonian}
\end{equation}
where $E_g$ is the band gap in GaAs. The second and the third terms describe, respectively, the motion of the exciton center of mass and the relative electron-hole motion. The last term describes the mixing of the relative motion and the exciton motion due to deviation of real crystal symmetry from the spherical one. The QW potential is not included in the Hamiltonian because of the large width of the QW. 
The effect of the QW interfaces will be discussed later, see Sect~\ref{sec:Wavefunction}, when the exciton wave functions will be analyzed.

The Schr\"{o}dinger equation with only the operator $\hat H_K$
determines the exciton motion in a crystal. The eigen states of this Hamiltonian can be represented as plane waves $\Psi_K(z) = \Psi_0\exp(i K z)$, where $\Psi_0$ is a normalizing constant, with the kinetic energy $E_{\text{kin}} = \hbar^2 K^2 / (2M)$. Here $M=m_{e}+m_{h}$ is a sum of electron and hole effective masses.

The operator $\hat H_{p}$ in Hamiltonian~(\ref{eq:Hamiltonian}) describing the relative electron-hole motion includes also the effect of electric field: 
\begin{equation}
\hat H_p=-\frac{\hbar^2}{2\mu}\sum_{\alpha=x,y,z}\frac{\partial^2}{\partial \alpha^2}-\frac{e^2}{\varepsilon_0r}+e(\mathbf{F\cdot r}).
\label{eq:Hamiltonian_p}
\end{equation}
Here $\alpha=\alpha_e-\alpha_h$, where $\alpha=x,y$, and $z$ denotes coordinates of the relative electron-hole motion, 
$\mu=m_em_h/(m_e+m_h)$ is the reduced exciton mass, and $e$ is the electron charge.
The Schr\"{o}dinger equation containing this operator will be used to obtain a wave function of the relative motion in Sect.~\ref{sec:Wavefunction}.

The last term in the exciton Hamiltonian~(\ref{eq:Hamiltonian}) plays an important role for the discussed phenomenon. 
It contains the cubic functions of the components of the free electron and hole wave vectors~\cite{IvchenkoPikus, PikusMaruschak, Cardona-PRB1988}:
\begin{equation}
\hat V=\sum_{\alpha=x,y,z} \left[\hat \varkappa_\alpha^{(h)}\gamma_v \left(a_1\hat J_\alpha+a_2\hat J^3_\alpha\right) + \hat \varkappa_\alpha^{(e)} \gamma_c\hat \sigma_\alpha\right].
\label{eq:cube}
\end{equation}
Here $\hat \varkappa_\alpha^{(h)}$ and $\hat \varkappa_\alpha^{(e)}$ are defined by $\hat \varkappa_z=\hat k_z(\hat k_y^2-\hat k_x^2)$ and $\hat\varkappa_x$  and $\hat\varkappa_y$ can be obtained by the cyclic permutation of indices. Operators $\hat k_\alpha$ describe the components of the wave vector of free electron or hole. Quantities $\gamma_v$, $\gamma_c$ are the material constants, whose values for the GaAs crystal are: $\gamma_c=24.5$ eV$\cdot$\AA$^3$, $\gamma_v=-74$ eV$\cdot$\AA$^3$~\cite{PikusMaruschak}. Dimensionless factors $a_1=13/8$ and $a_2=-1/2$ are defined by the symmetry of the crystal lattice. Matrices $\hat \sigma_\alpha$ and $\hat J_\alpha$ describe the electron and hole spin degrees of freedom, respectively~\cite{IvchenkoPikus}. 

A transition from the operators $\hat \varkappa_\alpha^{(e)}$ and $\hat \varkappa_\alpha^{(h)}$ to the wave vector of the center-of-mass exciton motion, $K=K_z$, and the momentum of relative electron-hole motion, $\hat p_{\alpha}=-i\hbar\partial/\partial\alpha$, is carried out by a substitution of expressions
\begin{equation}
\begin{split}
& \hat k^{(e,h)}_z=\pm\frac1\hbar\hat p_z+\frac{m_{e,h}}{M} K,\\
& \hat k^{(e,h)}_{x,y}=\pm\frac1\hbar\hat p_{x,y}.\\
\end{split}
\label{eq:wavevector}
\end{equation}
into Eq.~(\ref{eq:cube}). It gives rise to many terms containing different combinations of $K$ and $\hat k^{(e,h)}_{\alpha}$ in the first and second powers. We consider here only the term $\hat V_K$, which includes the wave vector $K$ in the first power. This is the term, which is required for the description of the effect under discussion. As our analysis shows, other terms only slightly affect the exciton energy, obtained from the Schr\"{o}dinger equation containing the first three terms of Eq.~(\ref{eq:Hamiltonian}). 

The linear-in-$K$ part of the operator~(\ref{eq:cube}) reads:
\begin{equation}
 \hat V_K=\frac{1}{\hbar^2} K (\hat p_x^2-\hat p_y^2) \Bigl[\gamma_v\frac{m_h}{M}(a_1\hat J_z+a_2\hat J^3_z)+\frac{m_e}{M}\gamma_c\hat \sigma_z\Bigr].
\label{eq:linear}
\end{equation}
Here $\hat J_z$ and $\hat \sigma_z$ are the diagonal matrices of the hole angular momentum and of the Pauli matrix. We should note that the operators $\hat J_z$, $\hat J_z^3$, $\hat \sigma_z$ as well as the wave vector $K$ change their sign at the time inversion operation so that the operator $\hat V_K$ remains invariant with respect to this symmetry operation.

To describe the experimentally observed effect, we should calculate the matrix element of the operator~(\ref{eq:linear}) on the exciton wave function. In the calculation of the wave function, we consider only the relative electron-hole motion described by the Hamiltonian~(\ref{eq:Hamiltonian_p}).
It means that we ignore the mixing of the relative electron-hole motion and the exciton center-of-mass motion. Estimates show that the mixing weakly affects the wave function. At the same time, even this simplified Schr\"{o}dinger equation with the Hamiltonian~(\ref{eq:Hamiltonian_p}) is rather complex for obtaining the wave function as it is described in the next subsection.

The matrix elements of $\hat V_K$ calculated with the exciton wave function, $\phi(F, r)$, acquire the form:
\begin{equation}
V=\pm\lambda(F) \zeta K.
\label{eq:linear_final}
\end{equation}
The dependence of these matrix elements on the electric field is controlled by the factor $\lambda(F)$,
\begin{equation}
\lambda(F)=\frac{1}{\hbar^2}\langle \phi(F, r)|\hat  p_x^2-\hat p_y^2|\phi(F, r) \rangle=\frac{1}{\hbar^2}\left(p_x^2-p_y^2\right).
\label{eq:lambda}
\end{equation}
We should emphasize here that $\lambda(F) \ne 0$ only if the electric field has a nonzero in-plane component. Indeed, if $F_x = 0$, the cylindrical symmetry of the problem is preserved, therefore $p_x^2= p_y^2$ and $\lambda(F) = 0$ at any $F$.

The factor $\zeta$ in Eq.~(\ref{eq:linear_final}) has a form:
\begin{equation}
\zeta\equiv  \left[\gamma_v\frac{m_h}{M}(a_1 J_z+a_1 J^3_z)+\frac{m_e}{M}\gamma_c \sigma_z\right].
\label{eq:zeta}
\end{equation}
Here $J_z=3/2$, $J_z^3=27/8$, and $\sigma_z=1/2$ are the diagonal elements of respective matrices. Signs ``$\pm$'' in Eq.~(\ref{eq:linear_final}) describe exciton states with projections of their angular momenta, $S_z = 1$ and $S_z=-1$, on the quantization axis ($z$ axis). The physical origin of the dependence of $V$ on the projection is related to the splitting of the electron and hole spin subbands in the GaAs crystal because of the absence of the inversion symmetry~\cite{IvchenkoPikus, PikusMaruschak, Cardona-PRB1988}. 

Taking into account Eq.~(\ref{eq:linear_final}), we can finally rewrite the exciton energy in an electric field as:
\begin{equation}
H(F,K) = E_g - R + \frac{\hbar^2K^2}{2M} \pm \lambda(F) \zeta K.
\label{eq:zero}
\end{equation}
Here $R$ is the interaction energy of electron and hole in the exciton.

We should note here that the hole Hamiltonian described in Ref.~\cite{PikusMaruschak} includes also terms linear in the hole wave vector, $\hat k_\alpha^{(h)}$. They read: 
\begin{equation}
\hat V_1=\sum_{\alpha,\beta,\delta}\gamma_v^{(1)}\hat J_\alpha(\hat J_\beta^2-\hat J_\delta^2) \hat k_{\alpha}^{(h)},
\label{V1}
\end{equation}
where $\gamma_v^{(1)}$ is a constant, which value for GaAs is given in Ref.~\cite{PikusMaruschak}. Indexes $\alpha\ne\beta\ne\delta$ equal to $x,y,z$. Using  equations~(\ref{eq:wavevector}), we can separate perturbation $\hat V_1$ into two parts. One of them contains $\hat p_{\alpha}$. Their matrix element are zero because $\langle \phi(F, r)|\hat p_{x,y,z}|\phi(F, r)\rangle=0$. Another one reads:
\begin{equation}
\hat V_{1z}=\gamma_v^{(1)}\hat J_z(\hat J_x^2-\hat J_y^2) \frac{m_h}{M} K.
\label{V1z}
\end{equation}
This perturbation mixes the heavy-hole and light-hole exciton states giving rise to their energy shifts,
\begin{equation}
E_{hh,lh}=\frac12\left[H_h+H_l\pm\sqrt{(H_h - H_l)^2 + 4V^2_{1z}} \right],
\label{eq:roots}
\end{equation}
where $V_{1z}=\gamma_v^{(1)}(\sqrt{3}/2)(m_h/M)K$.
These energy shifts do not depend on the direction of the exciton propagation and on the electric field. Thus, the linear in the hole wave vector terms do not contribute to the effect of phase inversion of the spectral oscillations.

\subsection{The wave function of the relative electron-hole motion in an electric field}
\label{sec:Wavefunction}

In order to calculate the factor $\lambda(F)$ we need to find the wave function $\phi(F, r)$ of the relative electron-hole motion in the exciton. This function should be the solution of a Schr\"{o}dinger equation with the Hamiltonian~(\ref{eq:Hamiltonian_p}) that contains an applied electric field.  
A standard approach to the solution of this equation is a transition to parabolic coordinates by substitution $\xi=r+z'$, $\eta=r-z'$, and $\varphi=\tan^{-1}(y'/x')$. Here the axis $z'$ is directed along the electric field direction, which is tilted by a small angle to our $z$ direction. Axes $x'$ and $y'$ are orthogonal with respect to $z'$. 
The wave function $\phi(F, r)$ can be presented in this basis in the factorized form~\cite{Blossey1970, Blossey1971}:
\begin{equation}
\phi(F, r) = A f(\eta) g(\xi) e^{\pm i\mathrm{m}\varphi}.
\label{eq:funck_par}
\end{equation}
Here $A$ is the normalization constant. Functions $f(\eta)$ and $g(\xi)$ are the eigenfunctions of the Hamiltonian~(\ref{eq:Hamiltonian_p}) rewritten in parabolic coordinates. More specific, they are the solutions of the following equations obtained from the Schrodinger equation with the Hamiltonian~(\ref{eq:Hamiltonian_p}):
\begin{equation}
\begin{split}
&\frac{1}{\eta}\frac{d}{d\eta}\Bigl(\eta\frac{df(\eta)}{d\eta}\Bigr)+\Bigl(-\frac{\mathrm{m}^2}{4\eta^2}-\frac{\nu}{\eta}+\frac{\mu R}{2\hbar^2}-\frac{\mu e F\eta}{4\hbar^2}\Bigr)f(\eta)=0, \\
&\frac{1}{\xi}\frac{d}{d\xi}\Bigl(\xi\frac{dg(\xi)}{d\xi}\Bigr)+\Bigl(-\frac{\mathrm{m}^2}{4\xi^2}+\frac{{\nu'}}{\xi}+\frac{\mu R}{2\hbar^2}+\frac{\mu e F\xi}{4\hbar^2}\Bigr)g(\xi)=0.
\end{split}
\label{parabolic_a}
\end{equation}
In these equations, $R$ is the interaction energy of electron and hole in the exciton. In the absence of an electric field, it is the exciton Rydberg constant,  $R = R_b=\mu e^4/2\hbar^2\varepsilon^2$. The parameter $\nu$ is introduced in these equations to separate the variables and $\nu'=\nu+\mu e^2/(\varepsilon_0\hbar^2)$. The value of $\nu$ should be chosen manually so that the energies $R$ obtained as the eigenvalues of each equation in the system~(\ref{parabolic_a}) coincide with each other.
Hereafter, we take into account that the reflectivity spectra are formed by $s$-like excitons with $\mathrm{m}=0$, which efficiently interact with the light wave. Therefore, we omit terms with $\mathrm{m}$ in Eqs.~(\ref{parabolic_a}).

An exact analytical solution of the equations~(\ref{parabolic_a}) cannot be obtained. Their eigenvalues and eigenfunctions can be found numerically. This approach is based on the second-order finite-difference approximation of the derivatives on the equidistant grids over variables $\eta$ and $\xi$. The obtained system of linear equations is then solved by the iterative method~\cite{Samarskii, Khramtsov-JAP2016}. However, for the tilted electric field considered in our case, the problem is more complicated and it cannot be described just by Eqs.~(\ref{parabolic_a}). Namely, the QW interfaces are tilted relatively to the electric field direction and, therefore, they reduce the symmetry of the problem so that the parabolic coordinates cannot be used anymore. It seems that the numerical solution of the initial Schr\"{o}dinger equation with the Hamiltonian~(\ref{eq:Hamiltonian_p}) is the most appropriate in this case. However, only the three-dimensional Schr\"{o}dinger equation corresponding the cylindrical symmetry of the problem can be numerically solved at present~\cite{Khramtsov-JAP2016, Belov-PhysE2019}. The tilted electric field brakes the symmetry so that a Schr\"{o}dinger equation of greater dimension should be solved. Currently, this problem has not yet been addressed in the literature as far as we know.

In order to evaluate the effect of the tilted electric field we consider two limiting cases.
First, we study only the longitudinal electric field, where the electric field is directed along the structure growth axis (axis $z$).
In this case, the cylindrical symmetry of the problem is preserved. As a result, the effect of the phase inversion of spectral oscillations is absent [see Eq.~(\ref{eq:lambda}) and the respective text]. Nevertheless, the consideration of this case is necessary because the $z$-component of the applied electric field determines the exciton wave function profile across the QW layer. Furthermore, this component considerably affects the exciton-light coupling constant $\hbar \omega_{LT}$ in Eq.~(\ref{Eq1}).

At the second step, we qualitatively consider the general case of the tilted electric field, namely, taking into account the electric field component directed along the QW plane (e.g., axis $x$) which is perpendicular to the growth axis. In this case, not only the exciton wave function is changed, but also the phase inversion in the calculated spectra is modeled. There is, however, an evident problem with the boundary conditions in this case because the function $g(\xi)$ may be non-zero at the infinitely large distance. We consider the boundary conditions for this case in a subsequent section. As we show below, the calculations performed in these limiting cases allow us to qualitatively explain the observed effect. Moreover, the numerical results obtained with the use of reasonable values of free parameters well agree with the experiment.

\subsection{Longitudinal electric field}
\label{sec:Fz}

\begin{figure}[h] 
\includegraphics[width=.9\columnwidth]{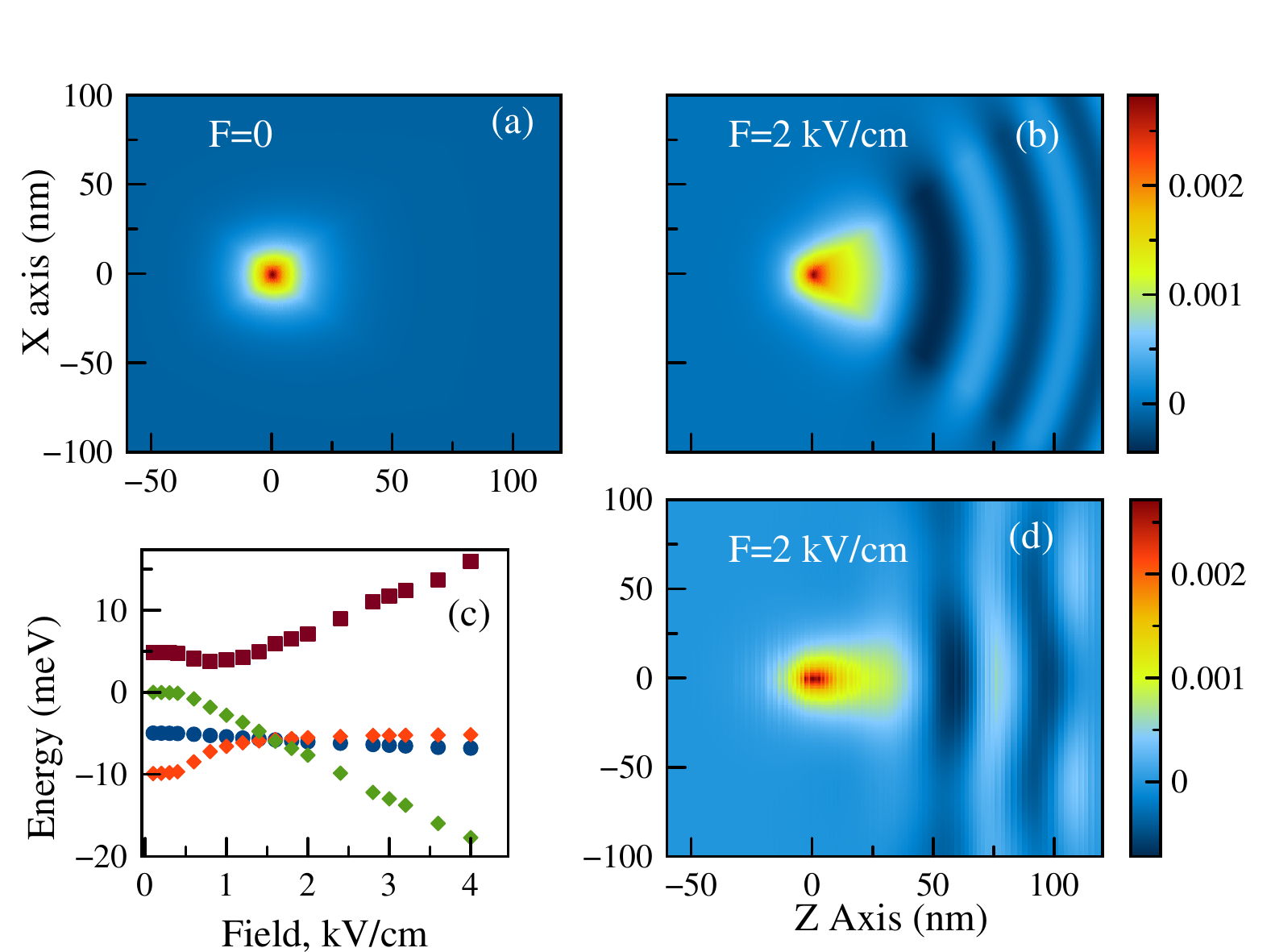}
\caption{(a), (b) Cross-sections of the wave function $\phi(F, r)$ in the $x$-$z$ basis calculated using Eqs.~(\ref{eq:funck_par}, \ref{parabolic_a}) in different electric fields $F = F_z$ given in each panel. (d) The cross-section of $\phi(F, r)$ obtained in the microscopic modeling (see the text for details). (c) The energy of the relative electron-hole motion in the external electric field (blue circles) calculated as the eigenvalue of the Hamiltonian~(\ref{eq:Hamiltonian_p}) and different contributions into this energy calculated as matrix elements of the specific Hamiltonian terms: black squares show the kinetic energy, red diamonds show the Coulomb energy, and green squares show the potential energy in the electric field.} 
\vspace{0.5cm}
\label{fig:wave-functions-z}
\end{figure}

Using equations~(\ref{parabolic_a}) we have numerically calculated the exciton wave function $\phi(F, r)$ for different values of the longitudinal electric field, $F_z$. 
Examples of the calculated wave functions $\phi(F, r)$ for zero electric field and for $F_z$ = 2 kV/cm are shown in panels (a) and (b) of Fig.~\ref{fig:wave-functions-z}. As one can see, the application of the electric field gives rise to the wave-like oscillations of $\phi(F, r)$. The oscillating behavior of the wave function is related to function $g(\xi)$ [see Eq.~(\ref{eq:funck_par})], which describes the spatial separation of the electron and the hole constituting the exciton. The oscillations of the function $g(\xi)$ resembles the oscillations of the Airy function describing the motion of a free carrier in an uniform electric field. The increase of the electron-hole distance is followed by the decrease of their Coulomb attraction and an increase of the exciton Bohr radius in the plane perpendicular to the $z$-axis. This increase is seen as a ``blurry'' of the wave function with increase of $z$, see Fig.~\ref{fig:wave-functions-z}(b).

To take into account the QW interfaces we assume that the maximum distance between the electron and the hole along the $z$-axis cannot exceed the QW width, $L_{QW} = 120$~nm. This condition partially determines the boundary conditions for the calculations. But, in the parabolic coordinates used in Eqs.~(\ref{parabolic_a}), this boundary condition is ambiguous. It correspond to a parabolic surface, rather than the flat one, where the exciton wave function acquires the zero value. Unfortunately, an attempt to write the boundary conditions for the flat interfaces inherent for QWs gives rise to the coupling of parabolic coordinates $\eta$ and $\xi$ and, correspondingly, the equations~(\ref{parabolic_a}). 

To accurately determine the exciton wave function in the presence of the real (flat) QW interfaces and of the electric field applied along the heterostructure growth axis, we have numerically solved a Schr\"{o}dinger equation for an exciton in a QW.  
The wave function has been calculated using the method described in Refs.~\cite{Khramtsov-JAP2016, Belov-PhysE2019}. This method is based on the finite-difference approximation of the three-dimensional Schr\"{o}dinger equation for the radial part of the exciton wave function. It allows us to accurately calculate  energies as well as corresponding wave functions of many quantum confined states in QWs of arbitrary widths. Here we should mention that, when a strong enough electric field is applied, the lowest energy state corresponds to the electron and hole states in the triangular potential wells, which much less interact with light.
To find the radiative exciton state, many low-energy states are calculated and a radiative constant, $\hbar\Gamma_0$, is determined for each of them~\cite{Ivchenko-book, Khramtsov-JAP2016, Khramtsov-PRB2019}. The state characterised by the highest value of $\hbar\Gamma_0$ is chosen as the bright exciton state in the electric field. In the numerical procedure, we assumed the QW barriers to be high enough to avoid tunneling of the exciton wave function beyond them. 
 
 The calculated wave function is shown in Fig.~\ref{fig:wave-functions-z}(d).
One can see the central peak due to Coulomb potential and the Airy-like behavior for $z>0$ where the QW potential becomes inclined deeper due to the electric field effect. With increase of the electric field strength, the number of Airy-like waves is also increasing. The numerically obtained function (microscopic model calculation) is similar to that calculated in the parabolic coordinates, though a noticeable difference is present. To evaluate the difference, we have calculated the radiative constant using both types of the functions. These constants coincide at zero electric field ($\hbar\Gamma_0 = 130$~$\mu$eV) and become different as the electric field increases. E.g., $\hbar\Gamma_0 (\text{micro}) = 70$~$\mu$eV and $\hbar\Gamma_0 (\text{parabolic}) = 71$~$\mu$eV for $F = 1$~kV/cm; $\hbar\Gamma_0 (\text{micro}) = 37$~$\mu$eV and $\hbar\Gamma_0 (\text{parabolic}) = 46$~$\mu$eV for $F = 2$~kV/cm. Nevertheless, this difference in not dramatic that allows us to use the wave functions obtained in the parabolic basis for a semi-quantitative analysis of the phenomenon under discussion.

Panel (c) in Fig.~\ref{fig:wave-functions-z} shows the electric field dependencies of various contributions to the total exciton energy described by the matrix elements of different terms in the Hamiltonian~(\ref{eq:Hamiltonian_p}). One can see that, the kinetic energy of the relative electron-hole motion described by the first term of the Hamiltonian~(\ref{eq:Hamiltonian_p})  slightly drops down at the electric field of about 1~kV/cm and then it almost linearly rises with the electric field. The drop is caused by the decrease (in an absolute value) of the electron-hole Coulomb interaction, $\langle \phi(F, r)|e^2/(\varepsilon_0 r)|\phi(F, r)\rangle$, due to the increase of the average distance between the carriers in the electric field. 

The linear rise of the kinetic energy at $F > 1$~kV/cm points out that the electron and the hole are accelerated almost as free carriers. Simultaneously, the potential energy, $\langle \phi(F, r)|e F z|\phi(F, r)\rangle$, of the carriers decreases in these electric fields so that their algebraic sum is almost zero. The total energy is mainly determined by the Coulomb energy, which remains nonzero even at large electric fields. This unexpected, at the first glance, effect is explained by the restriction of the carrier movement across the QW. Indeed, the electron-hole distance cannot be larger than the QW width, $L = 120$~nm. The calculations show that, at the fixed, large enough electric field, the Coulomb energy decreases as the QW width increases.

\subsection{Effect of the transverse electric field}
\label{sec:Fx}

\begin{figure}[h] 
\includegraphics[width=.8\columnwidth]{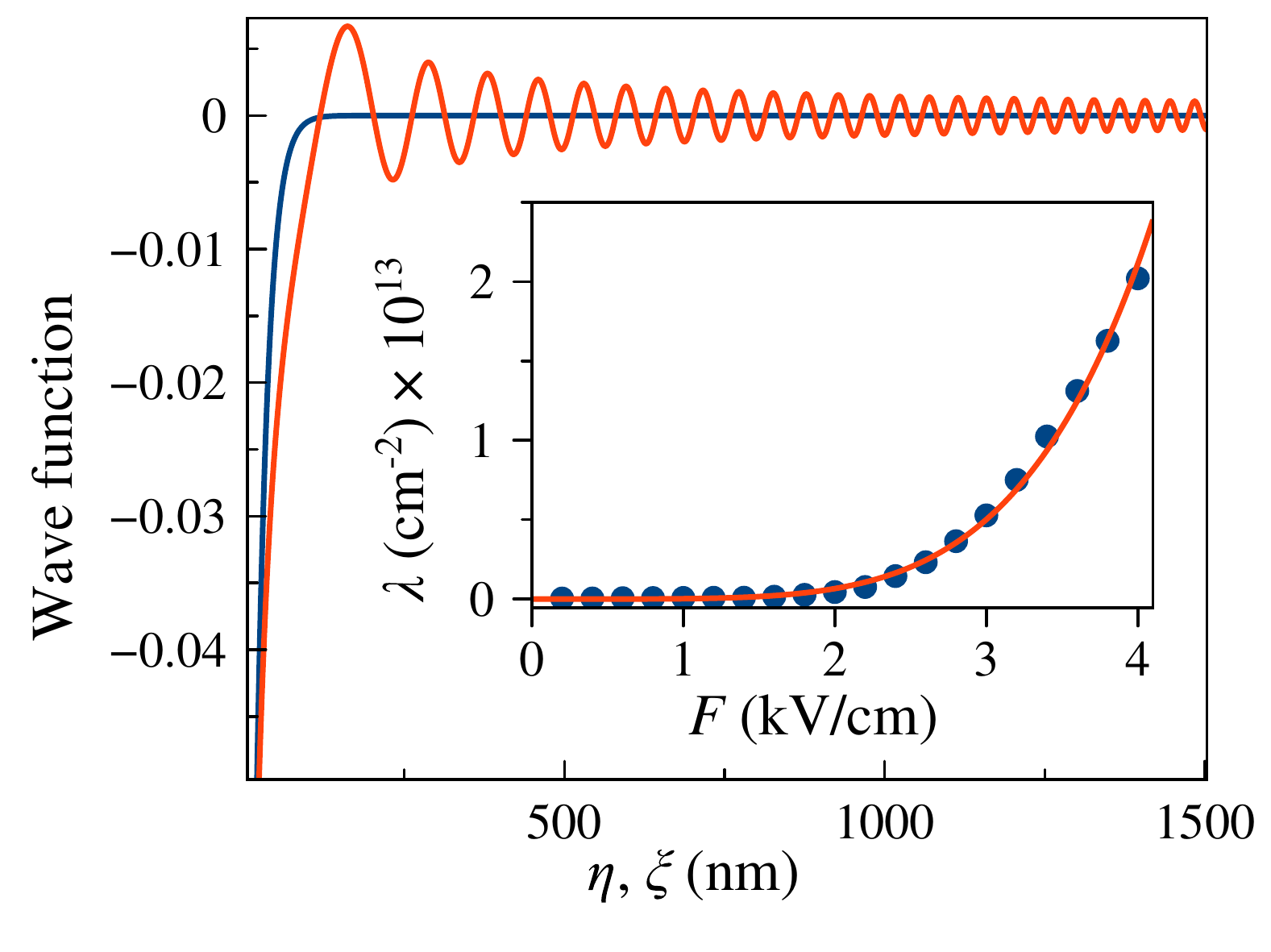}
\caption{Functions $f(\eta)$ (blue curve) and $g(\xi)$ (red curve) in an electric field $F_x = 1$~kV/cm. The inset shows the electric field dependence of the factor $\lambda(F)$ (points) calculated by means of Eq.~(\ref{eq:lambda}). The solid line is a phenomenological fit by the function $\lambda(F) = c F^5$, with $c = 2.1 \times 10^{10}$~(kV/cm)$^{-5}\,$cm$^{-2}$.} 
\vspace{0.5cm}
\label{fig:wave-functions-x}
\end{figure}

The transverse component of the electric field reduces the symmetry of the problem and makes it more complicated. Thus, we can discuss its effect only qualitatively. In the presence of the transverse component directed along the QW layer, say along the $x$ axis, there are no restrictions for the electron and the hole to run away from each other at the large enough field. In other words, there are, in principle, no zero boundary conditions along $x$ for the function $g(\xi)$ and, hence, for the wave function $\phi(F, r)$. An example of the functions $f(\eta)$ and $g(\xi)$ obtained by the numerical solution of Eqs.~(\ref{parabolic_a}) taking into account only the component $F_x$ is shown in Fig.~\ref{fig:wave-functions-x}. One can see that the function $f(\eta)$ rapidly decays to zero with the increase of the coordinate $\eta$. The function $g(\xi)$, however, noticeably oscillates even at $\xi$ close to 1500~nm. This gives rise to a problem of normalization of the function $\phi(F, r)$, in particular, finding of the coefficient $A$, see Eq.~(\ref{eq:funck_par}).

To solve this problem, we have to manually introduce the boundary conditions. For this purpose we take into account the finite exciton lifetime. For the excitons in the ground state, it is limited by the radiative decay time of several picoseconds in wide QWs~\cite{Khramtsov-JAP2016}. The application of an electric field partially suppresses the radiative decay and other processes, e.g., the exciton ionization and the phonon-mediated relaxation, limits the exciton lifetime, $\tau_X$, by approximately the same value. During the time $\tau_X$ the electron and the hole may tunnel through the Coulomb potential barrier inclined by the applied voltage and run away to some distance.
We evaluate the maximum distance between the electron and the hole considering a classical relation between the time and the distance for the uniformly accelerated carriers:
\begin{equation}
x_{\text{max}}=eF_x\tau_X^2/(2\mu).
\label{eq:delta_z}
\end{equation}

The exciton wave function is influenced by both components of the electric field that causes the function oscillations along both the $z$ and $x$ axes, see Figs.~\ref{fig:wave-functions-z} and \ref{fig:wave-functions-x}. The $z$-component of the applied electric field in the experiments under discussion is considerably stronger than the $x$ component. Therefore, we should take into account a strong modulation of the wave function along $z$ axis when analyzing its behavior along $x$ axis. A simple analysis shows that the modulation gives rise to the shrinkage of the wave function at small values of $z$, $|z| \le a_{\text{B}}$, where $a_{\text{B}}$ is the exciton Bohr radius. In other words, the probability to find the electron and the hole at the small distance $z$ is relatively high at any distance $x$ between them. Besides, the absolute value of the probability decreases in the presence of the longitudinal component by a factor $A_z = \omega_{\text{LT}}(F_z)/\omega_{\text{LT}}(0)$. The dependence of $\omega_{\text{LT}}(F_z)$ is discussed in the next section. We take into account these effects by normalizing the wave function to a quantity $A_z$ rather than to unity and consider this function only in the cylindrical volume limited by the $x_{\text{max}}$ along the $x$ axis and $\pm a_{\text{B}}$ along $z$. 

The limitation by the cylindrical volume leads to the increase of the amplitude of the oscillating ``tail'' of the wave function along the $x$ axis that results in the increasing factor $\lambda (F)$ calculated by means of Eq.~(\ref{eq:lambda}). In the calculations, we have taken $a_{\text{B}} = 15$~nm~\cite{Khramtsov-PRB2019} and $x_{\text{max}} = 1100$~nm at the electric field of $F = 4$~kV/cm. The latter quantity corresponds to the exciton lifetime $\tau_X \approx 2.5$~ps which appears to be reasonable in our case. The electric-field dependence of $\lambda(F)$ is shown in the inset of Fig.~\ref{fig:wave-functions-x}. As we show below, the magnitude of $\lambda(F)$ is sufficient to explain the observed phase inversion of the spectral oscillations.
   
\section{Modeling of the electroreflectance spectra}
\label{sec:spectrum}

As it is already discussed above (see Sect.~\ref{sec:theory}), a reflectance spectrum can be calculated in the framework of the model of polaritonic waves~\cite{Ivchenko-book, Kavokin-book2017,Tomasini-PRB1995, Loginov2006}. This model considers the exciton-like and photon-like polaritonic waves. Their dispersions are determined by equation~\cite{Ivchenko-book, Kavokin-book2017}:
\begin{equation}
\varepsilon(\omega,K)=\frac{c^2K^2}{\omega^2},
\label{eq:dispersion}
\end{equation}
where $c$ is the speed of light. This equation, together with the expression~(\ref{Eq1}), gives rise to a polynomial of the fourth order relative to the wave vector $K$. Its roots determine the dispersion dependencies for the polaritonic waves, $K_j(\omega)$. Two of these waves are the photon-like modes and the other two are the exciton-like ones. Two waves, one photon-like and one exciton-like, propagate in the forward direction and  two other waves propagate in the backward direction. 

The dispersions introduce the polarizability of the medium, $\chi(\omega, K_j)$, which is related to the dielectric function~(\ref{Eq1}) via the expression: 
\begin{equation}
\varepsilon(\omega,K_j)=\varepsilon_0+4\pi\chi( \omega, K_j).
\label{eq:eps}
\end{equation}

To find the amplitudes of polaritonic waves, the Pekar's additional boundary conditions (ABCs) are typically used~\cite{Ivchenko-book, Kavokin-book2017,Tomasini-PRB1995, Loginov2006}. They imply that the relation,
\begin{equation}
\sum_j\chi(\omega, K_j) E_j=0,
\label{ABC}
\end{equation}
is fulfilled at the QW interfaces. Here $E_j=E_j^{(0)}\exp[-i(K_jZ+\omega t)]$ is the electric field of the $j^{\text{th}}$ polaritonic wave and $K_j$ is its wave vector; $Z$ is the exciton center-of-mass coordinate. The condition~(\ref{ABC}) means that the total excitonic contribution into the polarizability at the QW boundaries should be zero. Besides, the standard Maxwell's boundary conditions (MBC) at the QW interfaces are used. These are the continuity of the tangential components of the electric field and of the magnetic induction of light waves.

The ABC and the MBC couple the amplitudes of the incident, transmitted, and reflected light waves with those of polaritonic waves in the QW. There are three equations for each interface that adds up to six equations in total. This system of equations is sufficient to find the amplitudes of four polaritonic waves inside the QW as well as the amplitudes of the transmitted and reflected light waves. The amplitude of the incident wave can be chosen arbitrary. The solution of this system allows one to find the ratio of the amplitudes of the reflected ($E_r$) and incident ($E_i$) light waves. Their ratio squared yields the reflection coefficient:
\begin{equation}
R(\omega, F)=\left|\frac{E_r}{E_i}\right|^2.
\label{R}
\end{equation}

\begin{figure}[h] 
\includegraphics[clip,width=0.6\columnwidth]{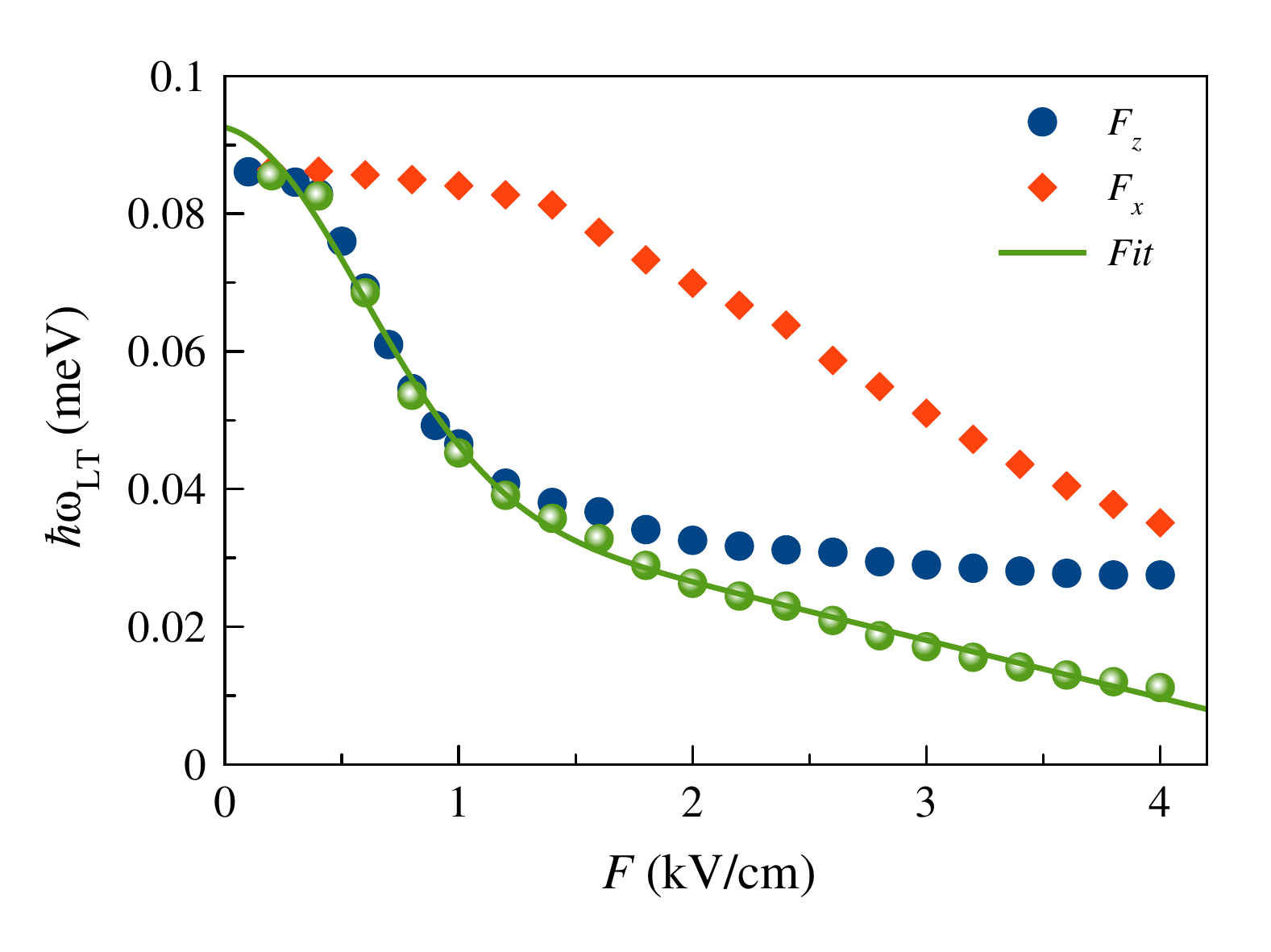}
\caption{Longitudinal-transverse splitting constant $\hbar\omega_{LT}$ as a function of the electric field applied across the QW layer ( $F_z=F\cos\alpha$, blue filled circles) and along it ($F_x=F\sin\alpha$, red diamonds); $\alpha = 20^{\circ}$.
Green balls show the effect of both component of the electric field calculated as: $\omega_{\text{LT}}(F) = [1/\omega_{\text{LT}}(0)]\, \omega_{\text{LT}}(F_z)\, \omega_{\text{LT}}(F_x)$. Solid curve is the fit by the phenomenological function: $f = 0.049 \exp{(-1.45 F^2)}-0.0084 F + 0.043$.
}
\vspace{0.5cm}
\label{fig:wLT}
\end{figure}

The developed theory allows us to model the reflectance spectra of the heterostructure with the QW using Eqs.~(\ref{Eq1}, \ref{eq:zero}, \ref{eq:dispersion}, \ref{eq:eps}, \ref{ABC}, \ref{R}). The modification of the spectra by the electric field is described by the factor $\lambda (F)$ shown in the inset of Fig.~\ref{fig:wave-functions-x}, as well as by the longitudinal-transverse splitting constant, $\hbar\omega_{LT}(F)$ entering in Eq.~(\ref{Eq1}).
The constant $\hbar\omega_{LT}(F)$ is calculated in a standard way~\cite{Ivchenko-book}:
\begin{equation}
\hbar\omega_{LT}=\Bigl(\frac{2eP_1}{E_g}\Bigr)^2\frac{\pi}{\varepsilon_0}|\phi(F, 0)|^2.
\label{eq:wLT}
\end{equation}
Here $\phi(F, 0)$ is the wave function of the relative electron-hole motion taken at the coinciding coordinates of the electron and the hole in the exciton; $P_1=\hbar p_{cv}/m_0$ where $p_{cv}$ is the interband matrix element of the electron momentum. For the GaAs crystal, $P_1=10.3\times10^{-5}$ meV$\cdot$ cm~\cite{PikusMaruschak}. Figure~\ref{fig:wLT} shows the dependencies of $\omega_{\text{LT}}(F)$ on the longitudinal and transverse components of the electric field. We have found that $F_z$ considerably affects the behavior of $\omega_{\text{LT}}(F)$. 
For example, for the field of $F_z = 2$~kV/cm, $\omega_{\text{LT}}(F)$ drops down to $0.37 \omega_{\text{LT}}(0)$. The effect of the transverse component of the electric field is noticeably weaker. It reduces $\omega_{\text{LT}}(F)$ only down to $0.87  \omega_{\text{LT}}(0)$ at $F = 2$~kV/cm.

In the calculations of the reflectance spectra, we have to introduce into Eq.~(\ref{Eq1}) a nonradiative broadening, $\hbar\Gamma = 2$~meV, to describe the experimentally observed relatively broad resonances, see Fig.~\ref{Fig1}. This broadening is probably caused by the electric-field-induced exciton ionization as well as by the fast energy relaxation of the excited quantum-confined exciton states with emission of phonons. Some additional broadening may be also caused by the generation of an electric current if the electric field is applied.

\begin{figure}[h] 
\includegraphics[clip,width=.60\columnwidth]{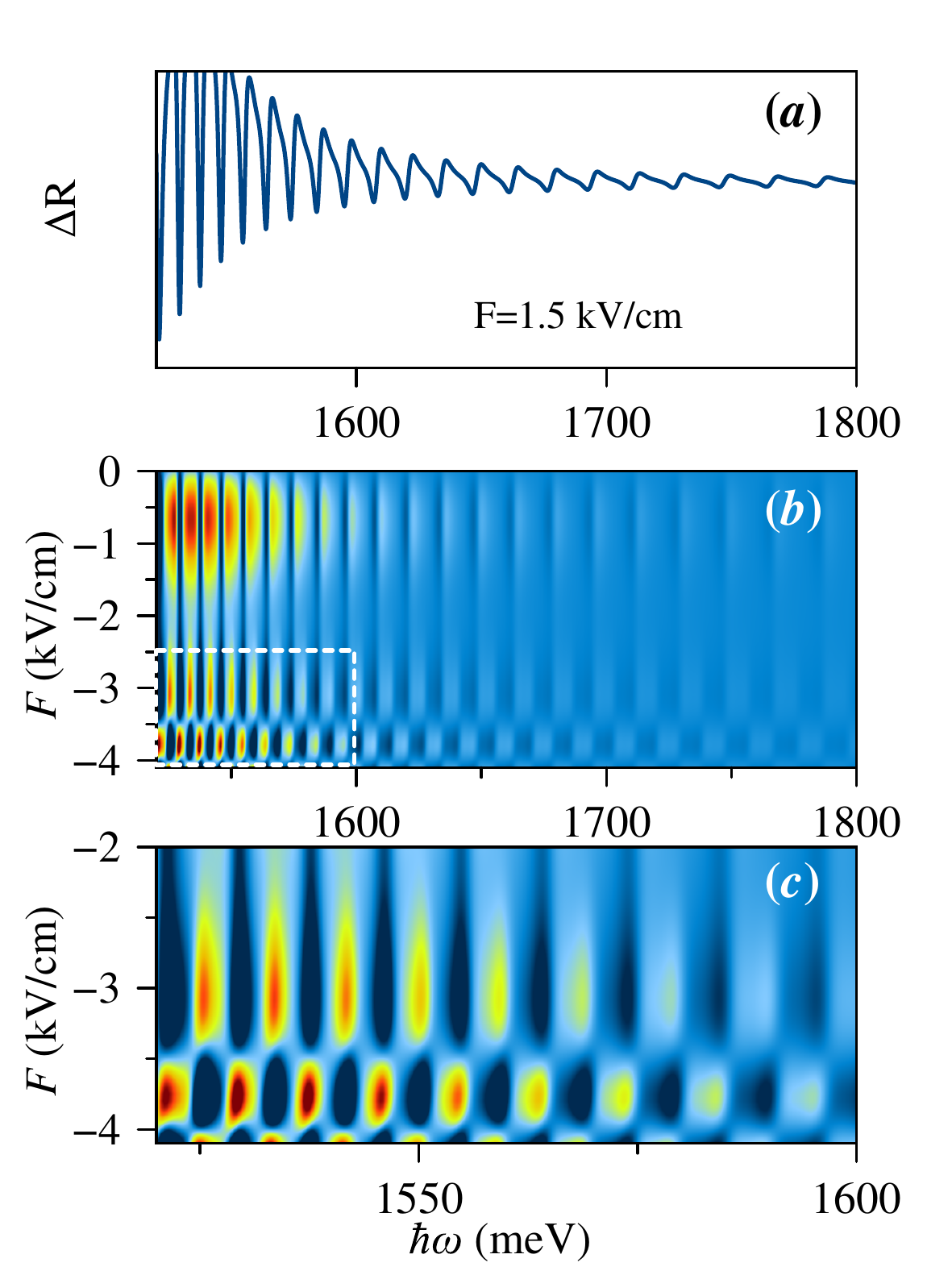}
\caption{ (a) A calculated differential reflectance spectrum at the electric field $F = 1.5$~kV/cm and (b) two-dimensional plot of the differential reflectance as a function of the photon energy and the applied electric field. Panel (c) shows an enlarged fragment of the plot demonstrating the phase inversion effect of the spectral oscillations.} 
\vspace{0.1cm}
\label{Fig5}
\end{figure}

For comparison with the experiment, the electroreflectance spectra, $\Delta R(\omega, F)$, are calculated using the numerical derivative of the total reflection $R(\omega, F)$ over $F$. The slowly varying background signal is subtracted. Figure~\ref{Fig5} shows the results of the calculations. One can see that the obtained dependence of $\Delta R(\omega, F)$ on the electric field well reproduces main features observed in the experiment (see Fig.~\ref{Fig1}). In particular, the phase inversion at $F = F_c \approx 3.5$~kV/cm is clearly seen in Fig.~\ref{Fig5}(c). \\

\section{Discussion}
\label{sect:discussion}

\begin{figure}[h] 
\includegraphics[clip,width=.70\columnwidth]{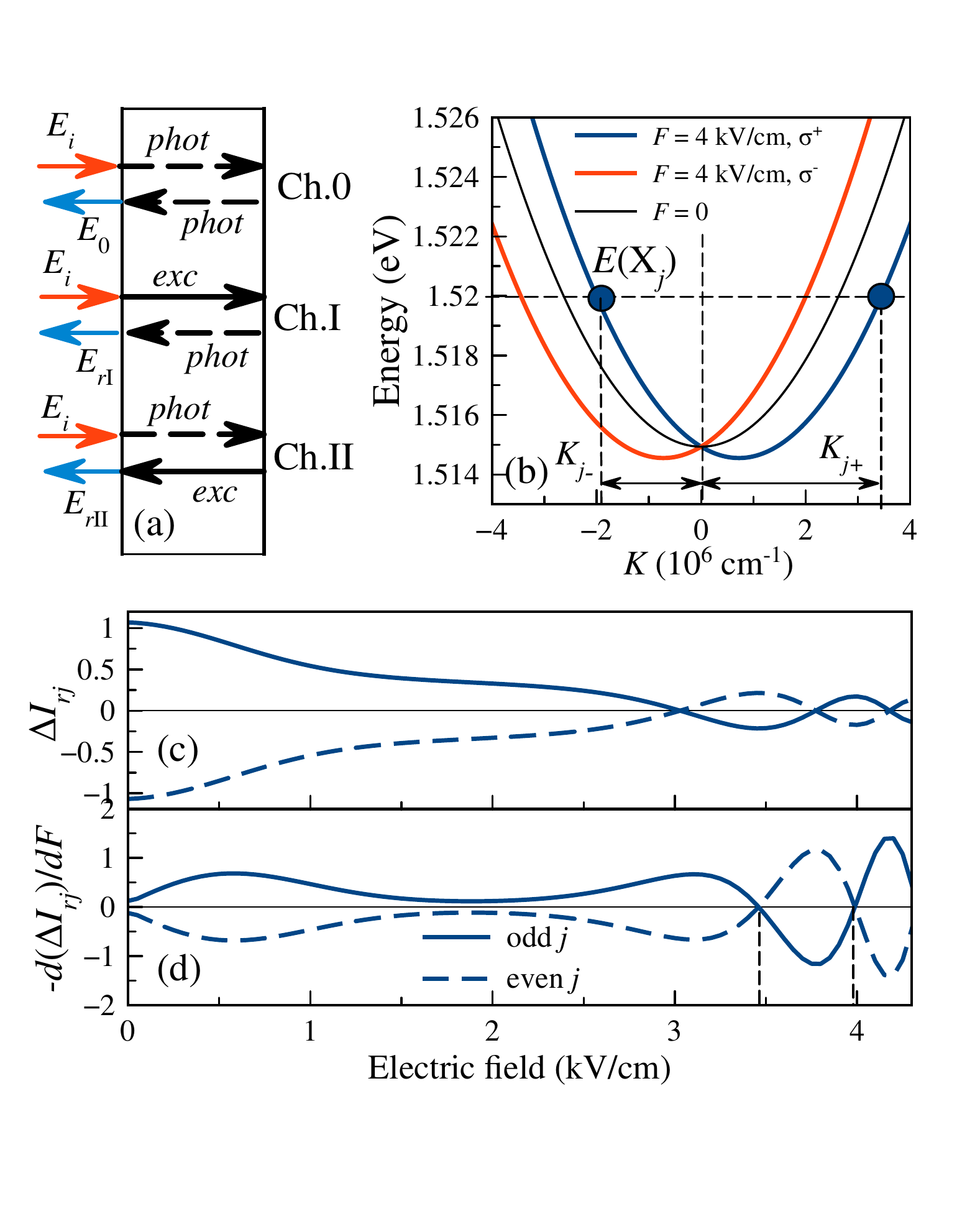}
\caption{(a) Schematic representation of the exciton-like and photon-like polariton modes in channels 0, I, and II. (b) dispersion dependencies for excitons at zero field (black curve) and field $F = 4$~kV/cm for the exciton angular momentum projections $S_z=+1$ (blue curve) and $S_z=-1$ (red curve). (c) The amplitudes of exciton resonances versus electric field for odd and ever number $j$ of the exciton states. (d) Derivatives of the amplitudes over $F$ multiplied by (-1) to match their sign with the experiment. } 
\vspace{0.1cm}
\label{Fig6}
\end{figure}

The microscopic origin of the phase inversion effect may be understood as follows. 
Let us assume that a light wave falls onto the QW from the left side [see Fig.~\ref{Fig6}(a)]. The polariton mode excited by the light wave in the QW is a composition of the photon-like and exciton-like modes. These modes are characterised by the wave vectors $K_p$ and $K_{\text{ex}}$, respectively. Their reflection from the right interface creates four waves because each mode can create both the exciton-like and photon-like modes. Thus, multiple polariton modes can propagate through the QW layer. They are discussed in Ref.~\cite{Loginov-PRB2014} in detail. 

At the QW interfaces, the polaritons are partially transformed into the outgoing light waves, which interfere with each other. The phases of these waves are determined by the number of passes of polaritons through the QW as the photon-like ($N_p$) and exciton-like ($N_{\text{exc}}$) modes, $\varphi = N_p \varphi_p + N_{\text{exc}} \varphi_{\text{exc}} = N_p K_p L + N_{\text{exc}} K_{\text{exc}} L$. The exciton and photon components of the polariton state are defined by the Hopfield coefficients~\cite{Hopfield-PR1958, Kavokin-book2017}.
For the polariton states, whose energies are far from the anti-crossing point of the exciton and photon dispersion curves, the fraction of light component is close to unity for the photon-like mode and to zero for the exciton-like one. The efficient conversion of the photon-like mode into the outgoing light wave leads to a relatively large background signal in reflectance spectra, which slowly varies with energy due to a small change of $K_p$ and, hence, of the phase $\varphi_p$. The main contribution to this signal comes from the one-time propagation of the photon-like mode in both directions. It is indicated as a Ch.0 in Fig.~\ref{Fig6}(a). 

The coupling of the exciton-like modes to the photonic continuum is less efficient and they are able to create relatively small peculiarities in the spectra. An analysis shows~\cite{Loginov-PRB2014} that the main contribution to the peculiarities comes from the polaritons, which propagate as an exciton-like mode in one direction and as a photon-like mode in the opposite direction. These channels of propagation are indicated as Ch.I and Ch.II in Fig.~\ref{Fig6}(a).

The constructive/destructive interference of the polaritonic waves propagating via channels Ch.0 and Ch.I, Ch.2 gives rise to the resonant peaks/dips observed in the spectra. 
In our case of the 120-nm QW, whose width approximately equals to a half of the light wave, the phase acquired by the polariton propagating via Ch.0, $\varphi_0 = 2 K_p L \approx 2\pi$. In this regime the constructive interference occurs for the exciton states with wave vector $K_{\text{exc}} = (2 N - 1) \pi /L$, where $N$ is a positive integer number. This is because the phase acquired in Ch.I and Ch.II, $\varphi_{I,II} \approx \pi + (2 N -1)\pi$, that is multiple of $2 \pi$. For the states with $K_{\text{exc}} = 2 N \pi /L$, the phase $\varphi_{I,II} \approx (2 N +1)\pi$, the interference is destructive, and the exciton states are observed as dips in the spectra.
For the high-energy exciton resonances where the parabolic dependence of the exciton dispersion is a good approximation, the condition of the constructive or destructive interference coincides with that of quantization of the exciton center-of-mass motion~\cite{Kiselev-pss1975, Kavokin-book2017}. 

In the tilted electric field, the linear-in-$K$ term in the exciton Hamiltonian~(\ref{eq:Hamiltonian}, \ref{eq:linear}) makes the propagation of the exciton-like modes in the forward and backward directions inequivalent. This term causes a constant shift by $\Delta K$ of the parabolic-like exciton dispersion along the $K$ axis, see Fig.~\ref{Fig6}(b). The value of $\Delta K$ is determined by the minimum of the exciton kinetic energy, $T=\hbar^2K^2/2M \pm \lambda(F) \zeta K$, see Eq.~(\ref{eq:zero}). Its derivative over $K$ gives rise to the required expression, $\Delta K(F)=\mp \lambda(F) \zeta M/\hbar^2$. The sign of this shift depends on the $z$-projection of the exciton angular momentum, $S_z$, see discussion after Eqs.~(\ref{eq:linear_final}). For excitons with $S_z = +1$ created by the $\sigma^+$-polarized light, the exciton dispersion curve is shifted towards the positive values of $K$ and, for $S_z = -1$ ($\sigma^-$-polarized excitation), this shift is negative, see blue and red curves in Fig.~\ref{Fig6}(b), respectively.

Let us consider a $j$-th quantum-confined exciton state with energy $E(\text{X}_j)$. For the right-hand shifted dispersion curve [blue curve in Fig.~\ref{Fig6}(b)], there are two types of excitons propagating in the forward and backward directions with effective vectors $K_{j\pm} = K_j \pm |\Delta K(F)|$, where $K_j = (\pi/L)j$. Correspondingly, the phase acquired by polaritons propagating via the I and II channels is:
\begin{equation}
\varphi_{j\pm} = \left(K_j \pm |\Delta K(F)|\right) L +\varphi_p.
\label{varphi}
\end{equation}
The last term in this expression, $\varphi_p = K_p L$, corresponds to the phase of the photon-like mode, which is predominantly independent of the electric field. The outgoing light waves, $E_{r\text{I}}$ and $E_{r\text{II}}$ [see Fig.~\ref{Fig6}(a)] acquire these phases similar to that of the incident light. The sum of these waves is proportional to:
\begin{equation}
E_{rj}(F) = 2 A_j \omega_{\text{LT}}(F) e^{i(K_j L +\varphi_p)} \cos[\Delta K(F) L].
\label{Erj}
\end{equation}
Here amplitude $A_j$ is determined by the amplitude of the incident light wave and by the coupling efficiency of the polariton waves to the continuum of the light waves~\cite{Loginov-PRB2014}.  The quantities depending on the electric field are explicitly specified in this expression. The light wave created by Ch.0 is: $E_0 = A_0 \exp(i 2\varphi_p)$.

The intensity of the reflected light detected in the experiments is governed by the sum of the light waves $E_0$ and $E_{rj}$:
\begin{eqnarray}
\label{Irj}
I_{rj} &=&  \left|E_0 + E_{rj}(F)\right|^2  
= A_0^2 + 4 A_0 A_j \, \omega_{\text{LT}}(F) \cos\left(\pi j +\varphi_p\right) \cos[\Delta K_j(F) L] \\ \nonumber 
&+& \left\{A_j \omega_{\text{LT}}(F) \cos[\Delta K(F) L]\right\}^2. 
\end{eqnarray}
The amplitude of the exciton resonances is much smaller than the background reflection amplitude. Therefore, we can neglect the last term in Eq.~(\ref{Irj}). 
Taking into account that $\varphi_p \approx \pi$, we finally obtain an expression for the amplitude of an exciton resonances:
\begin{equation}
\Delta I_{rj} = I_{rj} - A_0^2 \approx \pm 4A_0 A_j\, \omega_{\text{LT}}(F) \cos[\Delta K(F) L]. 
\label{DIrj}
\end{equation}

Figure~\ref{Fig6}(c) shows the electric field dependence of $\Delta I_{rj}$ for even and odd exciton states. At the electric field of $F < 3.2$~kV/cm, the amplitude of exciton resonance is positive for the odd number $j$ of the exciton state and negative for the even number that reflects the constructive/destructive interference of the corresponding light waves. At larger fields, the amplitude oscillates with an increasing rate, which is due to the rapid increase of the factor $\lambda(F) \propto F^5$, see the inset in Fig~\ref{fig:wave-functions-x}. 

In the experiments, the derivative of the reflection amplitude over the electric field is detected. The corresponding field dependencies are shown in the panel (d) of Fig.~\ref{Fig6}. They are multiplied by $(-1)$ to match with the phase of the experimentally detected signal (see Fig~\ref{Fig1}). As seen in Fig.~\ref{Fig6}(d), there is a critical value of the electric field, $F_c \approx 3.5$~kV/cm, at which the amplitude of the even and odd oscillations change their sign. This is the phase inversion effect, demonstrated in Fig.~\ref{Fig5}.

Theoretically, multiple phase inversions could be possible with a further increase of the electric field. Figs.~\ref{Fig5} and \ref{Fig6} show the next critical field for the phase inversion at about 4~kV/cm. However, we could not observe it experimentally because an electric current through the heterostructure grows superlinearly at the large applied bias.

A similar analysis can be performed for polaritons with the angular momentum projection $S_z=-1$. The dispersion curve for these excitons is shifted in the electric field to the left side in the reciprocal space, see Fig.~\ref{Fig6}(b). Because of the symmetry of the problem, the phase inversion for this case should occur exactly at the same critical electric field. Thus, the phase inversion effect should be observed at any polarization of incident light, in particular, for the linearly polarized light, as it can be represented as a superposition of the $\sigma^+$- and $\sigma^-$-polarized components.

\section{Conclusion}
\label{sec:conclusion}

The experimental study of electroreflectance spectra of a semiconductor structure containing a wide GaAs QW in the presence of an electric field applied at some angle to the growth axis has revealed a new effect. We have observed that the spectral oscillations related to the quantum-confined exciton states invert their phase at some critical electric field strengths. This phenomenon is theoretically analyzed in the framework of the polaritonic model. The analysis has shown that the phase inversion effect is caused by the electric-field-induced shift of the exciton dispersion curves, which is, in turn, described by the linear-in-$K$ term in the exciton Hamiltonian. 
The theoretical modeling of the spectra allowed us to reproduce all the essential spectral features observed in the experiment. 

We note that the electric-field-induced switching of exciton-polariton interference from the constructive one to the desctructive one and return is an analog of the Datta-and-Das effects that is in the heart of the spin transistor proposal~\cite{DattaDas}. A similarity can also be found between the observed switching and the concept of the excitonic transistor based on the an Aharonov-Bohm ring~\cite{Shelykh2009}. One can expect that results of the current work may be important for designing an exciton-polariton topological insulators, such as one discussed in Ref.~\cite{Klembt2018}. Tuning the in-plane electric field, one should be able to observe a transition from a trivial to a topological insulator regime for excitons.

\section*{Acknowledgments}

This work is supported by the Russian Science Foundation, 
grant No. 19-72-20039.
The calculations were carried out using the facilities of the SPbU Resource Center ``Computational Center of SPbU''.

\end{document}